\documentclass[pre,twocolumn,showpacs,preprintnumbers,amsmath,amssymb]{revtex4}
\usepackage{dcolumn}
\usepackage{bm}
\usepackage{amsmath,graphicx}
\usepackage{color}

\begin{document}
\title{Growth of aqueous foam on flexible membranes}

\author{Hiroyuki Shima}
\email[Email-address:]{shima@eng.hokudai.ac.jp}
\affiliation{Department of Applied Physics, Graduate School of Engineering,
Hokkaido University, Sapporo 060-8628, Japan}

\date{\today}

%%%%%%%%%%%%%%%%%%%%%%%%%%%%%%%%%%%%%%%%%%%%
%
%  Abstract
%
%%%%%%%%%%%%%%%%%%%%%%%%%%%%%%%%%%%%%%%%%%%%

\begin{abstract}
In this paper, I study the coarsening dynamics of two-dimensional dry foam sandwiched by deformable membranes. The time-varying deformation of the confining membranes gives rise to a significant alteration in the evolution of polygonal cells of bubbles when compared to the case of rigid membranes. This alteration is attributed to the correlation between the rate of inter-cell gas transfer and temporal fluctuation in surface curvature within a cell domain. The existing material constants are referred to understand the utility of the correlation effect toward the artificial control of the coarsening dynamics.
\end{abstract}

% -----------------------------------------------------------------


%% -----
%% 82.70.-y 	Disperse systems; complex fluids
%% 82.70.Rr 	Aerosols and foams 
%%
%% -----
%% 02.40.-k 	Geometry, differential geometry, and topology (see also section 04 Relativity and gravitation)
%% 
%% -----
%% 47.57.-s 	Complex fluids and colloidal systems 
%% 47.57.Bc 	Foams and emulsions 
%% 
%% ------
%% 81.10.-h 	Methods of crystal growth; physics of crystal growth 
%% 81.10.Aj 	Theory and models of crystal growth; physics of crystal growth, crystal morphology, and orientation 

%\keywords{ }

\maketitle

%------------------------------------------------------
\section{Introduction}

Aqueous foam exhibits a good interplay between geometry and physics. 
With time, foam consisting of polyhedral bubbles evolves into 
the equilibrium structure, during which internal gas diffuses 
from a bubble to others through thin curved liquid interfaces 
\cite{Weaire_1984,Glazier_1992,Stavans_1993,Vaz_2008}.
Diffusion is driven by the pressure difference between bubbles; 
assuming the constant diffusion coefficient, the pressure difference 
in two adjacent bubbles is proportional to the geometric curvature of 
their common boundary interface. 
Each boundary moves toward its concave sides due to the inter-bubble 
gas transfer, where the velocity of the boundary motion is again 
proportional to curvature \cite{Durian_1991}.
As a result, some bubbles dilate while others shrink and eventually disappear,
which results in a progressive increase in the average bubble size, 
{\it i.e.,} the coarsening of foam.

Foam that we encounter in our daily life, such as shaving cream and beer head, 
consists of a three-dimensional agglomerate.
Its coarsening dynamics \cite{Rivier_1983,Glazier_1993,Sire_1994,Monnereau_1998,Monnereau_2000,Hilgenfeldt_2001,
Cox_2003,Hilgenfeldt_2004,Glicksman_2005,Cantat_2005,Morawiec_2006,Thomas_2006,Kim_2006,MacPherson_2007,
Wang_2008,Gittings_2008}
as well as equilibrium cellular structures
\cite{Cox_2004,Kraynik_2004,Kraynik_2006}
have required formidable efforts for clarification because of geometrical and topological complexity.
This is partly the reason why a large degree of attention has been paid to two-dimensional counterparts,
{\it i.e.,} a foam monolayer confined between two 
membranes \cite{Weaier_1984,Glazier_1987,Glazier_1989,Stavans_1989,Stavans_1990,Tam_1996,Chae_1997,
Elias_1999,Graner_2000,Rutenberg_2006}.
It is remarkable that in two-dimensional foam confined between flat planes,
the time evolution of a polygonal cell of bubble depends only on the number of its sides,
regardless of its shape or area.
The growth-rate of the area $S$ of an $n$-sided cell is given by
$dS/dt \propto (n-6)$ \cite{vonNeumann_1952}.
This formula states that
a cell is stationary if $n=6$ but it grows (shrinks) at a constant rate if $n$ is
larger (smaller) than 6.
The disappearance of shrinking cells causes a topological change in the network of liquid interfaces,
whose effects on the stability of foam has also been largely investigated
\cite{Rutenberg_2006,Durand_2006,Cox_2008,Raven_2009}.

The coarsening behavior on the flat plane alters drastically when the foam is 
constrained to a curved surface \cite{Avron_1992,Peczak_1993}.
In the latter case, the evolution of cells is characterized by
the Gaussian curvature $K$ of the underlying surface.
When the surface has a positive (negative) curvature,
$n$-sided cells with $n<6$ ($n>6$) can be stationary, yielding $dS/dt = 0$,
only if $S$ equals to a specific value that depends on $n$ and $K$ (see Eq.~(\ref{eq_04n}) below).
Furthermore, the stability of those stationary cells is sensitive to
the sign of $K$.
For instance, no cell on a positively curved surface is stable;
once a cell grows (shrinks) slightly under perturbation, 
then it keeps growing (shrinking).
In contrast, all stationary cells on a negatively curved surface
are stable; therefore, the equilibrium configuration consists of various
$n$-sided cells each having a specific area determined by $n$($>6$) and $K$.
Such two-dimensional foam spreading over a curved surface could be realized
on an elastic confining plate or on a phase boundary with another fluid medium
that repels the foam.

The present article provides a further generalization of the coarsening
on curved surfaces, {\it i.e.},
the coarsening of foam constrained on a flexible surface exhibiting time-varying deformation.
The successive deformation of the confining surface changes the pressure difference
of adjacent cells that drives gas transfer across liquid interfaces.
As a consequence of the correlation between surface deformation and inter-cell gas transfer, 
the growth rate of the cell shows an intrinsic difference from that of the rigid curved surface. 
Realistic material parameters are employed to prove that the correlation effect plays 
a dominant role in the actual coarsening dynamics on flexible membranes 
under appropriate physical conditions.

%------------------------------------------------------
\section{Coarsening on a rigid curved membrane}

This section briefly reviews the coarsening theory of two-dimensional dry foam
on a rigid curved surface.
The term ``dry" refers to the assumption that liquid films between adjacent cells 
are so thin that they can be treated as curves with no thickness
and the vertices can be treated as points.
This assumption has long succeeded in exploring the nature of coarsening dynamics 
of foam, while considering that the realistic shapes of films with finite thickness 
\cite{Terraic_2006,Eri_2007,Grassia_2008,Marchalot_2009}
and their effects on permiability \cite{Lorenceau_2009}
may encourage quantitative agreements with experiments;
we shall revisit this point in Section V.

Let us assume that the monolayer foam is confined between two rigid membranes
with spatially uniform Gaussian curvature $K$.
The gap between the membranes is smaller than the typical length of boundary curves.
The growth-rate of the area of an $n$-sided cell with internal pressure $p$
is described by 
\begin{equation}
\frac{dS}{dt} = - \gamma \sum_{j=1}^n \Delta p_j \ell_j,
\label{eq_01}
\end{equation}
where $\Delta p_j \equiv p - p_j$ is the pressure difference between the cell and its
$j$th neighbor, $\ell_j$ is the length of the $j$th boundary curve separating the two cells, 
and $\gamma>0$ is a diffusion constant. 
Equation (\ref{eq_01}) captures the simple idea that if the cell has a 
higher pressure than the $j$th neighbor ({\it i.e.,} $p > p_j$),
then gas escapes to the neighbor, and vice versa.
With local equilibrium, $\Delta p_j$ is balanced by the line tension $\sigma$
along the interfaces,
satisfying the generalized Laplace-Young law \cite{Avron_1992}
\begin{equation}
\Delta p_j = \sigma \kappa_j,
\end{equation}
where $\kappa_j$ is the geodesic curvature of the $j$th interface.
From viewpoints of differential geometry, 
any $n$-sided polygon on a surface with curvature $K$
satisfies Gauss-Bonnet's theorem \cite{Millman_1977,Kamien_2002} expressed by
\begin{equation}
\sum_{i=1}^n (\pi - \alpha_i) + \sum_{j=1}^n \kappa_j \ell_j + \int \!\!\! \int K dS = 2\pi,
\end{equation}
where $\alpha_i$ is the internal angle of the $i$th vertex and must be
equal to $2\pi/3$ for all $i$
according to Plateau's lemma \cite{Plateau}.
Consequently, we obtain the result 
\begin{equation}
\frac{dS}{dt} = \gamma \sigma \left[ \frac{\pi}{3}(n-6) + K S \right],
\label{eq_03}
\end{equation}
which is called a generalized von-Neumann formula describing the foam coarsening 
on a rigid membrane \cite{Avron_1992}.

Formula (\ref{eq_03}) accounts for the stability properties of $n$-sided cells 
on rigid curved surfaces.
For $K>0$, such cells that satisfy $n<6$ and 
\begin{equation}
S = S^* \equiv \left| \frac{\pi}{K} \left( 2 - \frac{n}{3} \right) \right|
\label{eq_04n}
\end{equation}
can be stationary, although all stationary cells are unstable.
For instance, if $S$ becomes slightly larger than $S^*$ due to perturbation,
then the quantity in the square brackets in Eq.~(\ref{eq_03}) becomes positive. 
Therefore, we obtain $dS/dt > 0$ after the perturbation, 
which signifies a persistent growth in the cell.
On the contrary, all stationary cells are stable for $K<0$,
since $S$ being larger (smaller) than $S^*$ makes $dS/dt$ negative (positive).
In this context, the case of a flat plane
is marginal, in which the stationary cell of $n=6$ is
neither stable nor unstable against perturbation.

%------------------------------------------------------
\section{Coarsening on a deformable membrane}

Now, we focus our attention on the case where the confining membranes are mechanically flexible.
Membrane deformation induces a change in the value of $K$
within a cell domain,
thereby changing its area by incrementing $\delta S_{\rm cur}$.
In addition, inter-cell gas transfer contributes to the change in the area
by incrementing $\delta S_{\rm dif}$, which is similar to the case of rigid curved membranes
discussed in Section II.
An important consequence of membrane deformation is that 
it alters the internal pressure of the cell,
which causes a change in the pressure difference between cells
that works as a driving force to yield $\delta S_{\rm dif}$.
As a result, the membrane deformation and inter-cell gas transfer
correlate with each other, which imply a sizeable deviation in 
the growth rate equation from Eq.~(\ref{eq_03}) that we previously derived.
The quantitative determination of such a correlation effect on the cell growth equation
is the main purpose of the present work.

Let us assume that the Gaussian curvature $K$ of the confining membrane
is spatially uniform and varies continuously from $K=K_0$ at $t=0$
to $K_1 = K_0 + \delta K$ at $t = \delta t$.
The increment $\delta S$ of the cell area obtained at $t=\delta t$ is 
the sum of the two contributions:
\begin{equation}
\delta S = \delta S_{\rm dif} + \delta S_{\rm cur},
\label{eq_05c}
\end{equation}
where
\begin{equation}
\delta S_{\rm dif} = \left. \frac{dS_{\rm dif}}{dt} \right|_0 \delta t,
\label{eq_05x}
\end{equation}
and
\begin{equation}
\delta S_{\rm cur} = \left. \frac{d S_{\rm cur}}{d K} \right|_{K_0} \delta K,
\label{eq_05y}
\end{equation}
with $\delta K = (dK/dt) \delta t$.
The subscripts $0$ and $K_0$ imply differentiations
at $t=0$ and $K=K_0$, respectively.
$dS_{\rm dif}/dt|_0$ in Eq.~(\ref{eq_05x}) is identified with $dS/dt$ appearing in Eq.~(\ref{eq_03}),
since both describe the growth rate before the deformation occurs.
Equation (\ref{eq_05y}) has an explicit form that is given as
\begin{equation}
\delta S_{\rm cur} =
\frac{S_0^2}{2 \left(2\pi - K_0 S_0 \right)}
\delta K,
\label{eq_13x}
\end{equation}
and it shall be proved in Appendix A.
From Eqs.~(\ref{eq_05c})--(\ref{eq_13x}),
the area $S_1 = S_0 + \delta S$ at $t= \delta t$ is readily evaluated.

Next, we consider the area growth within the time duration $[\delta t, 2\delta t]$.
The increment $\delta S_{\rm cur}$ in this interval is obtained by replacing $K_0$ and $S_0$
in Eq.~(\ref{eq_13x}) by $K_1$ and $S_1$, respectively.
On the other hand, some caution is required in deriving the form of $\delta S_{\rm dif}$ at this stage
because of the correlation between internal pressure and membrane deformation.
The pressure $p$ of a given cell after deformation
is expressed by
\begin{equation}
p_1 = p_0 + \delta p_{\rm dif} + \delta p_{\rm cur},
\label{eq_09p}
\end{equation}
where $\delta p_{\rm dif}$ is the pressure increment that would be obtained
provided $\delta K = 0$,
and $\delta p_{\rm cur}$ is the one provided no diffusion occurs during deformation.
We denote by $p^j$ and $S^j$ the counterparts of the $j$th neighbor,
both of which obey the similar expressions of $p$ and $S$, {\it i.e.},
\begin{equation}
p_1^j = p_0^j + \delta p_{\rm dif}^j + \delta p_{\rm cur}^j,
\quad
S_1^j = S_0^j + \delta S_{\rm dif}^j + \delta S_{\rm cur}^j.
\label{eq_09q}
\end{equation}
To analyze the deformation effect on the pressure difference $\Delta p_1^j = p_1 - p_1^j$,
we take notice of the fact that the number of gas molecules in the cell
is preserved during deformation if no diffusion occurs.
This conservation law is formally represented by
$(p_0 + \delta p_{\rm cur})(S_0 + \delta S_{\rm cur}) = p_0 S_0$,
or equivalently
\begin{equation}
p_0 \cdot \delta S_{\rm cur} \;+\; \delta p_{\rm cur} \cdot S_0 \simeq 0,
\label{eq_10}
\end{equation}
which correlates $\delta p_{\rm cur}$ to $\delta S_{\rm cur}$.
We also see from the Laplace-Young law that
\begin{eqnarray}
& & p_0 - p_0^j = \sigma \kappa_0^j, \nonumber \\
& &(p_0 + \delta p_{\rm dif}) - (p_0^j + \delta p_{\rm dif}^j) 
= \sigma \kappa^j_{\delta t},
\label{eq_10r}
\end{eqnarray}
where $\kappa_0^j$ is the geodesic curvatures
of the $j$th boundary observed at $t=0$,
and $\kappa^j_{\delta t}$ is the fictitious one that would be observed
if $\delta K = 0$ during $[0,\delta t]$.
From Eqs.~(\ref{eq_09p})--(\ref{eq_10r}), it follows that
\begin{equation}
\Delta p_1^j = \sigma \kappa^j_{\delta t}
+
\left[ 
\frac{\delta S_{\rm cur}}{S_0} p_0 
-
\frac{\delta S_{\rm cur}^j}{S_0^j} \left(p_0 -\sigma \kappa^j_0 \right) 
\right].
\label{eq_09a}
\end{equation}
Substituting Eq.~(\ref{eq_09a}) in Eq.~(\ref{eq_01}) and applying Gauss-Bonnet's theorem, 
we obtain
\begin{eqnarray}
& & \left. \frac{dS_{\rm dif}}{dt} \right|_{\delta t} =
\gamma \sigma \left[ \frac{\pi}{3}(n-6) + K_0 S_0 + \delta K S_0 + K_0 \delta S_{\rm dif} \right] \nonumber \\
& & -\gamma \frac{\delta S_{\rm cur}}{S_0} p_0 \sum_{j=1}^n \ell_j 
+ \gamma \sum_{j=1}^n \frac{\delta S_{\rm cur}^j}{S_0^j} \left(p_0 -\sigma \kappa^j_0 \right) \ell_j,
\label{eq_15}
\end{eqnarray}
where the second-order terms with respect to increments were neglected.
It is to be noted that all the increments in the right side
are those obtained in the previous interval $[0, \delta t]$.

Equation (\ref{eq_15}) determines the diffusion-induced increment
$\delta S_{\rm dif} = dS_{\rm dif}/dt |_{\delta t} \delta t$ obtained
at $t=2\delta t$.
The area growth for larger $t$ can be evaluated by successively applying the procedure
shown above.
After deducing $S_2$ at $t=2\delta t$, for instance,
we rewrite the set $\{S_2, K_2, S_1, K_1, p_1\}$ by $\{S_1, K_1, S_0, K_0, p_0\}$
to calculate $S_2$ again,
which provides the subsequent value of the area at $t=3\delta t$.
An explicit algorithm of pursuing the time-varying $p$ is given in Appendix B.

It should be emphasized that in Eq.~(\ref{eq_15}),
the deformation effect manifests in the product $\delta K S_0$
in the square brackets and the two summations with respect to $j$.
In particular, the presence of the last two summations indicates that
the diffusion-induced growth rate of a cell
becomes dependent on the local environment around the cell.
This situation is in contrast to the case of a rigid membrane described by
Eq.~(\ref{eq_03}), where the growth rate is determined
only by the properties of the cell itself.

%------------------------------------------------------
\section{Deformation effect estimation}

To estimate the deformation effect in Eq.~(\ref{eq_15}),
we replace the fractions $\delta S_{\rm cur}^j/S_0^j$ by its mean value
over $n$ adjacent cells
and the sum of edge lengths $\sum_j \ell_j$ by
the perimeter ${\cal L}$ of an effective circular domain whose area equals to 
the original polygonal cell area $S$.
It follows that ${\cal L}$ at $t=\delta t$ is represented by
\begin{equation}
{\cal L} = \sqrt{4\pi S_1 - K_1 (S_1)^2},
\end{equation}
which will be derived in Appendix A (see Eq.~(\ref{eq_app06})).
Straightforward calculation yields
\begin{eqnarray}
& & \left. \frac{dS_{\rm dif}}{dt} \right|_{\delta t} 
= 
\gamma \sigma \left[ \frac{\pi}{3}(n-6) + K_0 S_0 \right](1-\beta) \nonumber \\ [4pt]
& & \qquad +
\gamma \sigma \left( \delta K S_0 + K_0 \delta S_{\rm dif} \right) 
+ \gamma p_0 (\beta  - \alpha) {\cal L},
\label{eq_18}
\end{eqnarray}
where
\begin{equation}
\alpha = \frac{\delta S_{\rm cur}}{S_0}, \quad
\beta = \frac1n \sum_{j=1}^n \frac{\delta S_{\rm cur}^j}{S_0^j}.
\end{equation}

To proceed with the arguments, we consider sub-millimeter-scale bubbles of $S\sim 0.1$ mm$^2$
under slightly time-varying curvature of $\delta K \sim 10^{-3}$ mm$^{-2}$ per second with $K=1$ mm$^{-2}$ at $t=0$;
these conditions are in the realm of laboratory experiments.
Then, we have $\beta \sim \delta S_{\rm cur}^j/S_0 \ll 1$, as a result of which
Eq.~(\ref{eq_18}) is simplified as
\begin{eqnarray}
\left. \frac{dS_{\rm dif}}{dt} \right|_{\delta t} 
&=& 
\gamma \sigma \left[ \frac{\pi}{3}(n-6) + K(t) \left(S_0 + \delta S_{\rm dif} \right) \right] \nonumber \\
&+& \gamma p_0 (\beta  - \alpha) {\cal L}.
\label{eq_021}
\end{eqnarray}
The most important deviation of Eq.~(\ref{eq_021}) from Eq.~(\ref{eq_03}), {\it i.e.,} from the growth rate equation
for rigid membrane cases,
is the presence of the term $\gamma p_0 (\beta-\alpha) {\cal L}$.
In fact, this term relates $\delta S_{\rm cur}$ obtained in the previous time interval, say, $[0,\delta t]$,
to $\delta S_{\rm dif}$ obtained in the subsequent interval, $[\delta t, 2\delta t]$.
When the confining membranes are rigid, then this term vanishes 
since $\alpha \propto \delta S_{\rm cur} \propto \delta K = 0$,
and so does $\beta$.

The salient finding of Eq.~(\ref{eq_021}) is the fact that by referring 
realistic material constants such as $p_0 = 10^5$ Pa, $\sigma = 10^3$ N/m, 
and $\gamma = 10^{-9}$ m/(Pa$\cdot$sec) \cite{Princen_1967,Marchalot_2009},
we obtain
\begin{equation}
\gamma p_0 (\beta - \alpha) {\cal L} \sim 10^{-5} \; {\rm mm}^2{\rm /sec},
\label{eq_020co}
\end{equation}
and
\begin{equation}
\gamma \sigma = 10^{-6} \; {\rm mm}^2{\rm /sec}.
\end{equation}
Therefore, the correlation-related term given in (\ref{eq_020co})
may be larger than (or comparable to, at least)
the coefficient $\gamma \sigma$ under the present physical conditions.
This means that the term should be dominant in rate equation (\ref{eq_021}),
thus totally changing the time evolution of cells from the case of rigid membranes.
An example to take note of is a situation wherein $S_0$ and $\delta S_{\rm cur} > 0$ are sufficiently large
to make the term $-\gamma p_0 \alpha {\cal L}$ dominant in Eq.~(\ref{eq_021}).
In this case, we obtain $\delta S_{\rm dif} < 0$ regardless of the values of $n$ or $S_0$,
which may result in the cell being stationary, {\it i.e.,} $\delta S_{\rm dif} + \delta S_{\rm cur} = 0$.
Therefore, one could stabilize the cell of a certain area $S_0$
by imposing appropriate curvature increments $\delta K$
whose value mainly depends on $S_0$ and $\gamma$.
This stability condition is totally different from that of cells confined in rigid curved membranes
where neither $S_0$ nor $\gamma$ but the sign of $K$ and the value of $n$
are relevant.

%------------------------------------------------------
\section{Summary and Perspectives}

Our results are based on the assumptions that all cell boundaries
have invariant material constants $\gamma$ and $\sigma$,
and the shape of each $j$th boundary curve is described by a constant $\kappa_j$.
In realistic foam, these parameters
are determined by the nature of liquid films that retain three-dimensional geometry
across the gap containing the foam.
In other words, each cell boundary is not a truly one-dimensional curve,
but a three-dimensional film with finite thickness 
whose value varies spatially within the film \cite{Terraic_2006,Eri_2007,Grassia_2008,Marchalot_2009}.
Therefore, membrane deformation will induce spatial fluctuations of $\gamma$ and $\sigma$
over the foam and that of $\kappa_j$ in each $j$th film.
The consideration of these fluctuations may enhance the quantitative precision
of the coarsening theory we have developed.

It is interesting to point out that membrane deformation may cause
a flow of liquid through the films,
as analogous to foam drainage in response to gravity and capillarity
\cite{Hutzler_2000,Hilgenfeldt_2001,Vera_2002,SaintJalmes_2006,Feitosa_2008}.
In fact, the deformation of the confining membranes
leads to the rearrangement of the liquid film network
as well as cell configuration,
and thus, inducing pressure gradient in the liquid.
Recently, it was shown that fluid flow on a surface with time-varying surface curvature
exhibits three kinds of dynamical responses depending on geometric and material constants \cite{Arroyo_2009}.
In this context, we conjecture that deformation-induced fluid flow in the current system
behaves differently from the case of a rigid membrane,
which gives rise to three-cornered coupling of
surface deformation with gas transfer and associated fluid flow.
Further examination incorporated with the theory presented in Ref.~\cite{Arroyo_2009}
will shed light on the issue.

In conclusion, we have studied the effect of surface deformation
on the coarsening dynamics of foam constrained in a gap of flexible membranes.
The growth rate equation of cells has been formulated
by taking into account the correlation between surface deformation
and gas transfer between adjacent cells.
We have found that the correlation-related term should be dominant 
in the resulting rate equation under realistic conditions, 
implying the possibility of artificial control of the coarsening dynamics 
through the confining membrane deformation.
It is hoped that the physical picture considered in this article
can provide a starting point for the analysis of the coupling 
between the membrane deformation and coarsening
dynamics in two-dimensional cellular structures.

%----------------------------------------------------
\acknowledgments

Illuminating discussions with Kousuke Yakubo, Akira Shudo, and Satoshi Tanda
are greatly acknowledged.
This study was supported by a Grant-in-Aid for Scientific
Research from MEXT, Japan,
and by Executive Office of Research Strategy in Hokkaido University.

%----------------------------------------------------------------------------------
\begin{figure}[ttt]
\includegraphics[width=5.0cm]{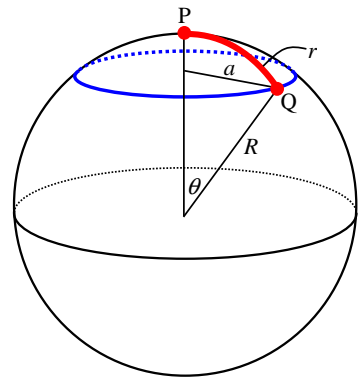}
\caption{Illustration for calculating the Gaussian curvature $K$ of a spherical surface with radius $R$.
The north pole $P$ is enclosed by a geodesic circle with radius $r$,
{\it i.e.,} by the locus of all points whose geodesic distances from $P$
equal to $r$.
The value of $K$ on $P$ is defined by Eq.~(\ref{eq_app01}).}
\label{fig_app1}
\end{figure}
%---------------------------------------------------------------------------------

%----------------------------------------------------
\appendix

%----------------------------------------------------
\section{Deformation-induced increment of cell area}

In this Appendix, we derive Eq.~(\ref{eq_13x}),
which is the expression of the deformation-induced increment $\delta S_{\rm cur}$
of a cell area.
For a general curved surface, the Gaussian curvature of a point on the surface is defined by
\begin{equation}
K = \lim_{r\to 0} \frac{3\left[ 2\pi r - {\cal L}(r) \right]}{\pi r^3},
\label{eq_app01}
\end{equation}
where $r$ is the radius of a geodesic circle around the point and ${\cal L}(r)$ is 
the lenght of its perimeter.
For a spherical surface with radius $R$, for instance,
it follows from Fig.~\ref{fig_app1} that ${\cal L}(r) = 2\pi a$, $a = R \sin \theta$ and $r = R \theta$.
Therefore, we obtain $K = 1/R^2 > 0$ and
\begin{equation}
{\cal L}(r) = \frac{2 \pi}{\sqrt{K}} \sin \left( \sqrt{K} r \right).
\label{eq_app02}
\end{equation}
Equation (\ref{eq_app02}) holds not only when $K>0$ but $K\le 0$,
as far as $K$ is constant with the circular region.

Now we consider the area $S$ of the circular region.
It is given by
$S = \int_0^r {\cal L}(r', K)  dr'$, and thus
\begin{equation}
S
= \frac{2\pi}{K} \left[ 1-\cos \left( \sqrt{K} r\right) \right].
\label{eq_05a}
\end{equation}
Eliminating $r$ from Eqs.~(\ref{eq_app02}) and (\ref{eq_05a}) yields
\begin{equation}
S(K,{\cal L}) = \frac{2\pi - \sqrt{4\pi^2 - K {\cal L}^2}}{K},
\label{eq_app06}
\end{equation}
which converges to ${\cal L}^2/(4\pi)$ in the limit of $K\to 0$.
Finally, we obtain
\begin{equation}
\left.\frac{\partial S}{\partial K}\right|_{{\cal L}}
= \;\;
-\frac{2\pi}{K^2}
+ \frac{\sqrt{4\pi^2 - K {\cal L}^2}}{K^2}
+\frac{{\cal L}^2}{2K \sqrt{4\pi^2 - K {\cal L}^2}},
\end{equation}
and equivalently,
\begin{equation}
\left.\frac{\partial S}{\partial K}\right|_{{\cal L}}
=
\frac{S^2}{2 (2\pi - KS)},
\end{equation}
which completes the proof.

%----------------------------------------------------
\section{Successive relation for internal pressure}

The internal pressure $p_{\mu}$ at $t= \mu \delta t$ $(\mu\ge 1)$ is given by the following procedure.
First, $\delta p_{\rm cur}$ is deduced from Eq.~(\ref{eq_10}) as
\begin{equation}
\delta p_{\rm cur} = -p_{\mu-1} \frac{\delta S_{\rm cur}}{S_{\mu-1}}.
\end{equation}
Next, $\delta p_{\rm dif}$ is derived by substituting Eq.~(\ref{eq_10r}) 
into the Gauss-Bonnet theorem,
which leads to
\begin{equation}
- \delta p_{\rm dif} {\cal L}_{\mu} 
+ \sum_{j=1}^n \left( p_0^j \delta \ell^j + \delta p_{\rm dif}^j \ell_1^j \right)
=
\sigma K_{\mu-1} \delta D_{\mu-1},
\label{eq_app_b4}
\end{equation}
where $\delta \ell^j = \ell_1^j - \ell_0^j$.
The summations in Eq.~(\ref{eq_app_b4}) are canceled out
if the sign of the summed terms is positive or negative depending on $j$.
As a result, we obtain
\begin{equation}
\delta p_{\rm dif} = - \sigma \frac{K_{\mu-1} \delta D_{\mu-1}}{{\cal L}_{\mu}},
\end{equation}
which gives $p_{\mu} = p_{\mu-1} + \delta p_{\rm dif} + \delta p_{\rm cur}$ for all $\mu\ge 1$
through the successive relations.

%%%%%%%%%%%%%%%%%%%%%%%%%%%%%%%%%%%%%%%%%%%%
%
%     References
%
%%%%%%%%%%%%%%%%%%%%%%%%%%%%%%%%%%%%%%%%%%%%

\end{document}